**A low power flexible dielectric barrier discharge disinfects surfaces and improves the action of hydrogen peroxide.**

Sophia Gershman[1*], Maria Belen Harreguy Alfonso[2], Shurik Yatom[1], Yevgeny Raitses[1], Phillip Efthimion[1], and Gal Haspel[2]

[1]Princeton Plasma Physics Laboratory, Princeton, NJ, USA
[2]New Jersey Institute of Technology, Newark, NJ, USA
*sgershma@pppl.gov

**Abstract**

There is an urgent need for disinfection and sterilization devices accessible to the public that can be fulfilled by innovative strategies for using cold atmospheric pressure plasmas. Here we explore and evaluate an approach combining a flexible printed circuit design of a dielectric barrier discharge (DBD) with an environmentally safe chemical reagent for surface decontamination from bacterial contaminants. The device operates in ambient air without any additional gas flow at a power density of about 0.5 W/cm$^2$. Using a 3% hydrogen peroxide solution as a model reagent we demonstrate improved reduction in the bacterial load of a surface contaminant of >6log$_{10}$ in 90 seconds, about 3log$_{10}$ and 2log$_{10}$ faster compared to hydrogen peroxide alone and the flexible DBD alone, respectively, for the same treatment time. The synergistic action of the plasma bioactive properties and hydrogen peroxide result in a dramatic improvement of surface disinfection. This opens new possibilities for using the low power flexible DBD plasma sources for surface disinfection and decontamination.

## I. Introduction

There is an urgent need for wide use of sanitizing and disinfecting agents and techniques. Brought into focus the current COVID-19 pandemic, it is no longer limited to medical, pharmaceutical, or food industry, but rather expanded to the decontamination of commonly used surfaces such as doorknobs and devices, such as masks, cell phones, and pens. Over the last two decades, cold atmospheric pressure plasmas (CAP) have seen rapid development in the areas of bacterial and viral inactivation and surface disinfection [1-8]. A recent review [8] summarizes the achievements of a broad range of CAP plasma sources, including dielectric barrier discharges (DBD), that effectively inactivate bacteria, viruses, fungi, and bacterial spores. In spite of these achievements, a practical CAP device is not available to the public. Some devices require vacuum, others use sophisticated power sources, compressed gases, or long treatment times that make them impractical for personal use. Here we investigate the first hand-held chemically augmented plasma device for rapid disinfection. The device is safe, flexible, and suitable for personal use. We demonstrate faster disinfection than plasma or chemicals alone in stable low power operation.

The disinfecting and even sterilization effectiveness of plasmas are due to their bio active properties such as UV, electric fields, and reactive nitrogen and oxygen species (RNS, ROS) [9-12]. The mechanisms of bacterial inactivation have been investigated by many groups but remain unclear. The chemical and electrical plasma properties may be affecting a bacterial cell in stages. The electrons and the electric field affect the cell membrane and aid in the cell penetration by the RNSs and some long-lived ROSs. ROS are involved in lipid peroxidation and other oxidative reactions damaging the cell membrane and aiding the transport of RNS/ROS into the cell. Inside the cell the ROS/RNS damage proteins, lipids, and the DNA. The combined effect of these processes is bacterial cell inactivation [8, 13].

Most of the work on medical and biological applications of DBDs has been conducted on one of three configurations, a one-electrode configuration, a plasma jet, and a less common surface DBD [8, 14-21]. In a one-electrode device the high voltage electrode is encased in a dielectric material and the treated surface acts as a ground electrode [8, 14, 15]; the treated surface is exposed to high electric fields and fluxes of charged particles. The most extensively studied is the plasma jet, which uses power from pulsed dc to microwave range and where plasma effluent is carried by a gas flow to the treated surface. The plasma effluent is suitable for medical applications but requires a compressed gas supply [14, 16-18]. Surface DBD has been primarily studied as an actuator for flow control in aeronautics applications and for large area surface modifications (ex. [19]). Introduction of devices based on flexible printed circuit design facilitated its applications in medical and biological fields [5, 20-22]. The device studied here is based on a flexible printed circuit design.

Atmospheric pressure plasmas have been shown to be effective for the decontamination of surfaces from bacteria and viruses, but the level and the rate of inactivation strongly depend on the biological species, experimental conditions, and the plasma source. For example, D-value (time for 1log10 reduction) is 225 s for the exposure to the gases produced by a DBD [23], 150 s for a DBD [2], 35 s for *E. coli* exposed to an atmospheric pressure helium/air glow discharge [3], and 15 s for a paper DBD [21]. The fast reduction D=15 s was achieved by a single-use flexible DBD device using a printed patterned electrode on a paper substrate and operated at 2 kHz, 3.5 kV AC, 10 W. This device burns out in under 2 m of operation. This diversity of results and conditions means that a new device such as ours has to be tested and the disinfection has to be demonstrated using standard microorganisms.

Another variation on plasma disinfection is the use of low-pressure plasma activated hydrogen peroxide vapor [13 24, 25]. Systems such as Sterlis® [26] are widely accepted methods of sterilization of materials susceptible to high temperatures, humidity, and corrosion. In more recent studies, the addition of hydrogen peroxide has been explored to enhance plasma disinfection at atmospheric pressure [4, 27-29]. 6log$_{10}$ reduction was achieved with the addition of H$_2$O$_2$ droplets into a corona discharge and greater than 6log$_{10}$ reduction in the bacterial load and a significant reduction in biofilm and spores by adding H$_2$O$_2$ vapor to the plasma effluent [27, 29].

Here we present a flex-DBD that is based on the printed circuit design [5, 21]. It is safe to touch as other DBD devices, but unlike a jet, it does not need any additional gas supply or sophisticated power sources, and is capable of long term stable operation unlike a paper DBD device. We demonstrate the fast disinfection effect in less than 90 s of this device on the standard bacteria, gram-negative *Escherichia coli* (*E. coli*). We also demonstrate the combined effect of 3% H$_2$O$_2$ commonly available antiseptic and the flex-DBD device for surface decontamination. The device tested here could be used to disinfect surfaces and personal items and protection equipment such as masks.

II. **Experimental methods ,procedures device assembly and characterization methods**

The flex-DBD is based on a printed circuit design [5, 21]. The flex-DBD consists of a layer of copper tape (16 mm x 26 mm) serving as a high voltage electrode and covered by a layer of Kapton® tape 100 μm thick ($\epsilon_{rel}$≈3.5), and a patterned ground electrode, ENIG® coated, 30 μm thick placed on top of the Kapton® tape (Fig. 1-a). The ground electrode has 200 (10x20) square cavities 0.75 mm x 0.75 mm.

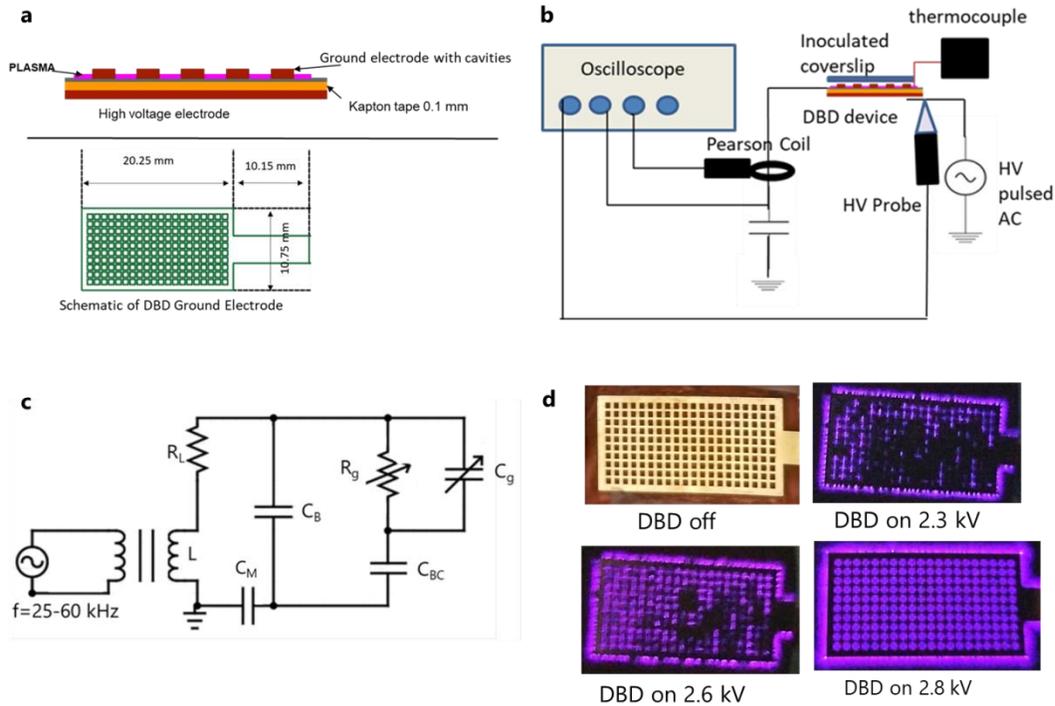

**Figure 1. The flex-DBD is based on a printed circuit design and operates at <3 kV, ~40 kHz, <0.5 W/cm² supplied by a portable power source.** (a) The design of flex-DBD consists of conducting and insulating layers and a patterned ground electrode. (b) The experimental setup includes a portable pulsed ac high voltage (HV) source and monitoring apparatus. The patterned electrode is grounded. (c) The equivalent circuit, where $C_B$ is the capacitance of the flex-DBD device excluding the rectangular cavities, $C_{BC}$ is the capacitance of the solid portion and $C_g$ and $R_g$ are the capacitance and resistance of the open air portion of the cavities, $C_M$ is the measurement capacitor, and L and $R_L$ are the inductance and the resistance of the secondary coil of the high voltage transformer of the high frequency power source. (d) As the applied voltage increases, a greater part of the surface of flex-DBD lights up.

A regulated, 500 V – 10 kV, 25-60 kHz, pulsed AC power source (PVM 500 AC, Information Unlimited) was used to generate the discharge (Fig. 1-b). The patterned electrode of the flex-DBD was connected to the power supply ground and the copper foil electrode is connected to the high voltage output transformer of the power supply (Fig. 1b and 1-c). A Tektronix D (2 GS/s, 250 MHz) oscilloscope was used for monitoring the current, voltage, and charge transfer in the circuit during the disinfection experiments. Current to ground was measured by a Pearson Model 2877 Current Monitor (1 V/A, 2 ns rise time). Tektronix P6015 HV probe was used for measuring the voltage at the high voltage (copper tape) electrode. The charge transferred was determined by measuring the voltage across a 10 nF capacitor (Fig. 1-b and 1-c, $C_M$) connected in series on the ground side of the flex-DBD.

The flex-DBD was operated in resonance mode; the parallel connection between the DBD and the secondary coil on the high voltage output transformer has a resonance frequency, $\approx \frac{1}{2\pi\sqrt{LC}}$, where L, is the inductance of the secondary (Fig. 1-c) and C is the capacitance of the flex-DBD. During operation the capacitance of the flex-DBD is approximately, $C \approx C_B + C_{BC}$. At the resonance frequency, the overall impedance of the L-C circuit as seen by the power source is at a maximum, therefore minimizing the current drawn and hence the power used by the device (Fig. 1-c). The voltage is maximized facilitating the discharge at lower power. To start the device, the voltage was adjusted to a value below the starting voltage, and the frequency was then adjusted achieve the maximum voltage. The voltage was then increased to the critical value of about 1900 V at which the discharge starts. Once the flex-DBD was lit, the frequency was adjusted to return to the resonance and the voltage was increased until the entire surface of the flex-DBD looked lit to the naked eye (Fig. 1-d). The resonance/operating frequency was (42 ± 2) kHz. The variation in the resonance frequency is likely due to the slight differences in the hand-made devices and the operating conditions. The temperature of the grounded glowing face of the flex-DBD was monitored for several minutes to ensure a steady state condition was maintained. Maximum applied voltage and duty cycle were adjusted to maintain the temperature <50 C in steady state operation. Except for the low power trial at 2 kV, the disinfection experiments were conducted with the peak voltage of 3 kV, the displacement current amplitude of 50 mA, and a duty cycle of ~20% and pulse repetition rate of 1 kHz. The voltage, current, and charge measurements were conducted during the sterilization experiments.

We used a Princeton Instruments PIMAX-3 ICCD camera for the fast imaging of the surface discharge. The camera was triggered on a high current spike and minimal delay was used so that the camera was open for the next current spike. The timing of the current spikes and camera shutter was monitored on the oscilloscope making the appropriate adjustments in the lengths of the connecting cables. Therefore, the ICCD images were correlated to the current spikes recorded by the Pearson coil.

**Sterilization efficiency experiments**

To demonstrate the disinfection ability of the flex-DBD, we tested the effectiveness of the device in reducing the bacterial load of *E. coli* 10-beta (New England Biolabs) and the Standard *E. coli* AMS 198 (ATCC-11229). Bacteria were cultured following vendor's instructions at 37°C in Luria broth or Luria Agar (both from Research Products International). For the surface test experiments, we used a suspension of the *E. coli* strain that expresses cytoplasmic green fluorescent protein (OP-50-GFP) and forms a uniform bacterial lawn rather than discrete colonies.

The flex-DBD was attached to a holder or in other experiments to the lid of a 60 mm petri dish (Fig. 2-a). The experiments included the treatment of bacteria seeded in petri dishes, on a disposable textile type material, on metal (aluminum), and on glass microscope cover slips. For treatment of bacterial plates, we spread 50 µl of a fresh bacteria culture on LB-Agar plates and treated the plate surface with the flex-DBD attached to a lid placed over the petri dish (Fig. 2-a) for different amounts of time. We then incubated the plates overnight at 37°C and visually compared them to untreated control and examined for areas that were clear of bacteria. To test the disinfection of the textile-type surfaces, 100 µl of *E. coli* OP-50 was spread on Tyvek®. The petri dish cover with the flex-DBD attached was placed over the inoculated area. The petri dish cover used in this experiment was cut to maintain a 1-2 mm distance between the treated surface and the face of the flex-DBD. Each region was treated for a set amount of time and at the end of the treatment time, immediately stamped with an LB contact plate (Carolina Biological Supply Company). The untreated area was stamped as the control. The contact plates were incubated for 24 hours at 37°C. Qualitative results were assessed visually by observing GFP expression using an epifluorescence imaging station (FastGene Blue/Green LED GelPic Box, Nippon Genetics).

To quantify disinfection, bacteria (*E. coli* 10-beta, Standard *E. coli* AMS 198 or OP50-GFP) were inoculated on glass microscope coverslips, 25 mm in diameter. Four droplets of 5 µl each (20 µl) of the bacterial culture (starting concentration $10^8$ CFU/ml) were placed onto each coverslip and allowed to dry for approximately 40 minutes. The slides with dry bacterial culture were then placed onto the flex-DBD with the inoculated side directly in contact with the discharge (Fig. 2-b). The coverslips were treated for 10 s, 30 s, 90 s, 270 s. At the end of the treatment time, the treated coverslip was placed in a centrifuge tube with 7.5 ml of Luria broth, enough to cover the coverslip. The tubes were vortexed on a medium setting for 20 seconds to recover the bacteria from the treated surface but not damage the cell membrane. The resulting bacterial suspension was plated on LB agar and the plates were incubated for 24 hours. Cultures were then counted and the number used to calculate the logarithmic reductions of bacterial concentration. All disinfection experiments were conducted with the flex-DBD operating at 3 kV, 20 % duty cycle, and 40 – 50°C.

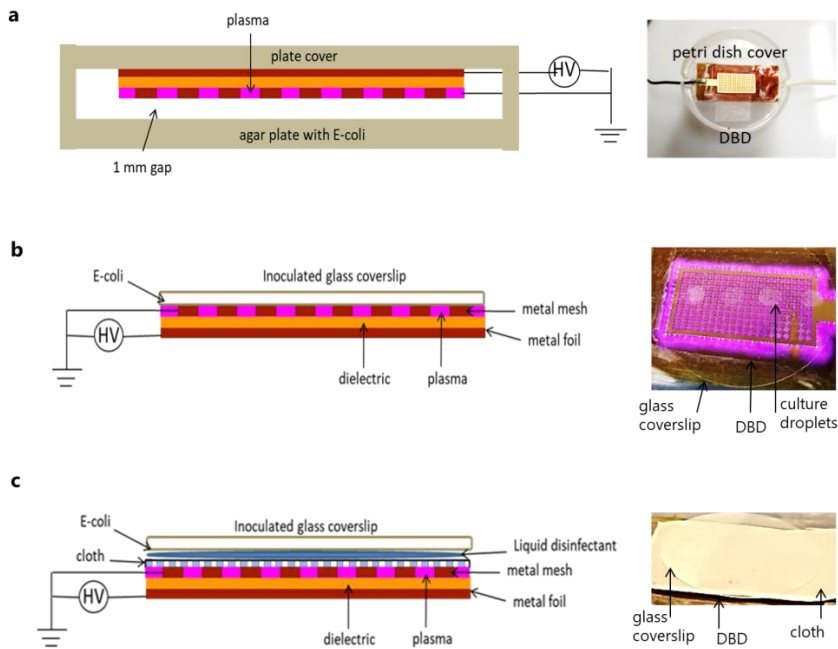

**Figure 2. The flex-DBD can be used a short distance away, in contact with a surface, and with a liquid disinfectant.** (a) The flex-DBD treatment of a petri dish: flex-DBD is fixed to the inside of a petri dish lid. (b) Contact treatment: inoculated coverslip is directly in contact with the flex-DBD and (c) Combined treatment with the flex-DBD and $H_2O_2$: flex-DBD is covered by a semi-permeable cloth with a liquid disinfectant on top; the inoculated cover slip is on top, with the bacterial culture in contact with of the 3% $H_2O_2$ (20 µl dried culture, 10 µl 3% $H_2O_2$).

We tested flex-DBD disinfection efficiency in conjunction with a commonly available 3% solution of $H_2O_2$. The discharge in the flex-DBD is suppressed by water so we used a semipermeable material (Tyvek®) to keep the water from the $H_2O_2$ solution from the surface of the flex-DBD (Fig. 2-c). We disinfected pieces of Tyvek by soaking for 5 minutes in a 70% solution of isopropyl alcohol, then dried thoroughly for at least 30 min. For each trial, we placed a piece of the sterile material on top of the operating flex-DBD and five 2 µl droplets of 3% $H_2O_2$ water solution were placed on top of the Tyvek®. An inoculated glass coverslip was then placed on top of the solution with the bacteria in direct contact with the solution (Fig. 2-c). At the end of each treatment, we dropped both the coverslip and the cloth into a centrifuge tube with 7.5 ml of LB solution. The same recovery and plating procedure was used in all experiments. The controls for this experiment were untreated coverslips as well as treating the bacteria with $H_2O_2$ alone (the same procedure except for that the flex-DBD stayed turned off).

We tested the disinfection efficiency of $H_2O_2$ aided by the UV light produced by the flex-DBD. We placed a thin film filter transparent down to 190 nm between the DBD and $H_2O_2$ to block all the output from the plasma except light.

We used two indicator tests to check the $H_2O_2$ production by the plasma. A 2 – 200 ppm range indicator was used to test the production of $H_2O_2$ in ≈20 µl of Luria Broth and a 1 – 10% range indicator was used to check the changes in the concentration of in the 3% $H_2O_2$ solution used for the disinfection that combined the flex-DBD and $H_2O_2$.

To eliminate the temperature as a factor contributing to the disinfection process, we placed inoculated glass coverslips on a heating block at 47°C and repeated the same procedures as for the disinfection to determine the reduction in the bacterial load. We did not observe any reduction in the number of CFU/ml.

**Disinfection data analysis**

To quantify bacterial load reduction, we imaged the treated plates after 24 h of growth and counted the number of bacterial colonies, interpreted as colony forming units (CFUs) in the plate-seeding solution. When possible ImageJ (1.x) software [33] was used to count CFUs, otherwise we counted visually. The bacterial load in CFU/ml was calculated by dividing the CFU count by 40 µl (the volume inoculated on each plate) and multiplying to account for the serial dilutions and by 375 (sampling 20 µl from coverslips washing volume of 7.5 ml LB broth). The reduction in the bacterial load was calculated as $log_{10}(\frac{N_o}{N})$, where $N_o$ is the bacterial load without any treatment (0 s coverslip), and N corresponds to each treatment time. All the experiments were performed in triplicates of samples and plates. Statistical significance between pairs of treatments was evaluated using paired ANOVA.

III. RESULTS AND DISCUSSION

**Surface dielectric barrier discharge in a flex-DBD device**

A 40 kHz sinusoidal voltage is applied to the high voltage electrode for ~200 µs (20% duty cycle at 1 kHz repetition rate). A surface dielectric barrier discharge ignites inside the cavities and around the perimeter of the flex-DBD (Fig. 1a and 1d). The discharge propagates along the surface of the dielectric and eventually erodes the substrate of the ground electrode. The erosion pattern observed on used devices indicates that the discharge occurs in the center portion of each cavity leaving the angles intact, and effect that is due to the geometry of the electric field. The erosion of the substrate of the patterned electrode happens over months of operation possibly because only a few cells appear lit at on a fine time scale. This is seen from the ICCD images synchronized with the current spikes (Fig. 3-b).

Water can form a conductive film on a surface which prevents charge accumulation on the electrode and as a result breakdown conditions cannot be reached [21, 36-38]. This effect depends on the voltage rise time. At sub-nanosecond rise times, the breakdown voltage may be reached, but in our case, the voltage rise time is ≈7 µs, too slow to prevent the charge leakage from the conductors. The parts of the device that become moist will not light until dry. When the power is on, heating occurs due to the resistive and dielectric losses and the device will self-dry and restart once the moisture evaporates. In the experiments with $H_2O_2$ solutions, a water resistant material is used to prevent the flex-DBD becoming wet but to allow the active species from the plasma to reach the treated surface. This sensitivity to water is important for bio-related applications.

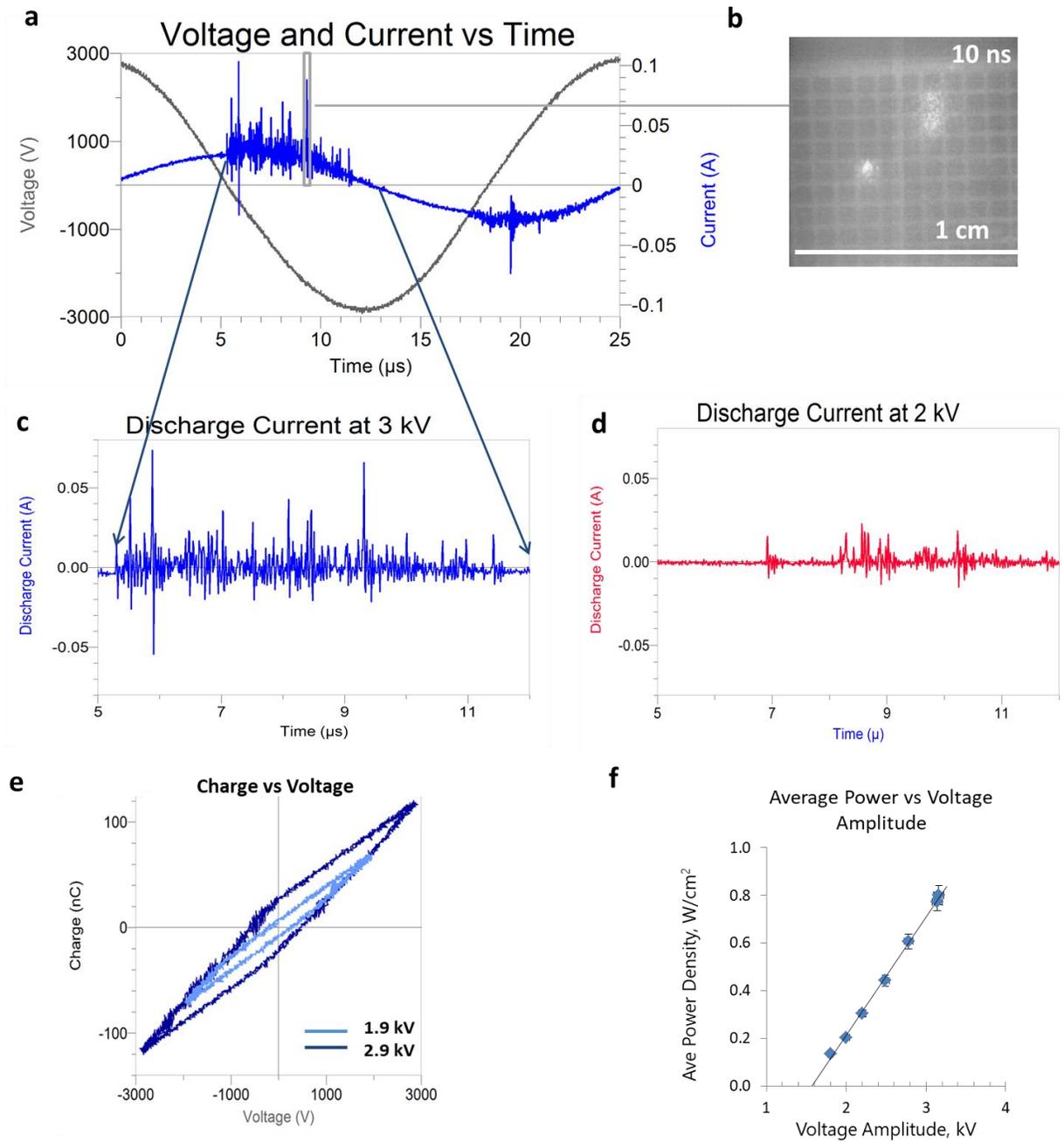

**Figure 3. Electrical characteristics of the flex-DBD.** (a) Current and voltage traces for one ac cycle. (b) A 10 ns exposure ICCD image synchronized with an individual current pulse. (c) The discharge current for one quarter of the cycle for 3 kV and (d) for 2 kV applied voltage amplitude. (e) The Lissajous figures for one full cycle of the sinusoidal applied voltage of an amplitude of 1.9 kV and 2.9 kV. The 1.9 kV trial had 0.05 mJ/cycle, average discharge power of 0.3 W, and the 2.9 kV – 0.14 mJ/cycle, 0.75 average discharge power. (f) The average power as a function of the applied voltage amplitude. The duty cycle and the frequency were kept constant at 20% and 40 kHz respectively.

A typical current trace is comprised of a displacement current sinusoidal component of (42 ± 2) kHz and the superimposed sharp spikes 10 – 50 ns in duration corresponding to the discharges (Fig. 3-a, 3-c, 3-d). The displacement current was subtracted from the total measured current to obtain the discharge current (Fig. 3-c). The number of discharges, their overall duration, and their amplitude increase with increasing voltage (Fig. 3-c and 3-d). Although at 3 kV, the flex-DBD appeared completely lit, fast imaging triggered on a current spike demonstrates that during each current spike only a few bright regions are observed (Fig. 3-b). Individual current spikes appear to correspond to isolated discharge events that appear randomly on the surface of the DBD. The number and the amplitude of the current spikes is not symmetrical with a greater number of spikes corresponding the mesh electrode acting as the anode. In case of a positive mesh electrode (anode), the electrons are able to flow into the anode and the current grows, but if the mesh electrode is negative, the electrons accumulate on the dielectric and the current stops resulting in lower current spikes. This asymmetry has been observed in plasma actuators that are a single edge surface DBD similar to the flex-DBD [19]. The number of individual discharges or current peaks varies depending on the maximum applied voltage (overvoltage). For example, Fig. 3d (Fig. 3-c) the number of current peaks (over 10 mA) is ≈15±8 at 1.9 kV and increases to ≈45±8 for 3 kV (Fig. 3-c). The greater number of current spikes results in a greater amount of charge transferred in the circuit as evident from the Lissajous plots (Fig. 3-e).

The Lissajous plot has a two-slope shape with a slight asymmetry due to a greater number of more intense discharges for the negative voltage (positive patterned electrode). The energy dissipated in the circuit per one cycle can be calculated as the area of the Lissajous plot, and the power is then determined using the frequency, $f$, the duty cycle, $v$,

$$P = fv \int Q dV, \qquad (1)$$

Where Q is the charge measured by the capacitor probe and dV is the voltage obtained by the high voltage probe. For example, for the peak voltage of 1.9 kV the energy per cycle was 0.04 mJ/cycle. For the frequency of 41 kHz and a 20% duty cycle this gives the power of 0.3 W. For the max voltage of 2.9 kV the energy per cycle was 0.14 mJ/cycle, and the power, 1.1 W. The corresponding power density for the ~2 cm$^2$ device is 0.15 – 0.5 W/cm$^2$. The applied max AC voltage was varied from 1.6 kV to about 3 kV while keeping the frequency and the duty cycle constant. The resulting power varied linearly (Fig. 3-f) with the applied voltage, which can be used as a calibration curve to set the desired power for a given device.

Increasing the operating voltage increases the discharge power and corresponds to an increase in the number of individual discharges and the production of plasma. Increasing the duty cycle increases the overall power consumption by the device, but does not change the number of individual discharges.

**Disinfection using the flex-DBD**

To evaluate the effectiveness of the flex-DBD device in decontamination of surfaces from biological contaminants, we conducted qualitative and quantitative experiments. The qualitative experiments included the treatment of *E. coli* in petri dishes, the decontamination of inoculated aluminum and fabric surfaces (Fig. 4a and 4-b); the quantitative bacterial load reduction was determined by treating bacterial culture dried onto glass coverslips (Fig. 4-c and 4-d).

We placed droplets of the bacterial culture on an aluminum surface, treated by exposure to the flex-DBD, and then stamped with contact plates. The flex-DBD effectively reduced the bacterial load when the DBD was placed 1 mm from the surface and operated at 3 kV and 44 kHz, duty cycle of 20%. The flex-DBD device was also effective at disinfecting textile-type (Tyvek®) material. To assess to spatial extent of disinfection we uniformly inoculated the fabric with *E. coli* that was modified to express green fluorescent protein (OP-50-GFP) and treated a 10x20 mm area. The operating parameters remained the same, but the flex-DBD was placed directly onto the surface. There was a marked reduction in GFP positive colonies around the treated area (Fig. 4-b). Only viable bacteria contain GFP and fluoresce when excited with blue light. The distance of the DBD from the surface is also important because reducing the distance from 1 mm above the surface to a direct contact with the ground electrode, increased the rate of inactivation of bacteria. We obtained a similar spatial patter by treating *E. coli* bacterial culture dried onto the surface of a glass coverslip. A dark region, about the size of the DBD indicate *E. coli* that are not fluorescent because they did not survive the plasma treatment (Fig. 4-b), following 30 s treatment with the flex-DBD.

To quantify the bactericidal effect of the flex-DBD we inoculated and dried glass microscope slides and measured the surviving bacterial load in colony forming units per milliliter (CFU/ml). Treatment with the flex-DBD device reduced viable bacteria $\log_{10}(N_o/N)$= 4.1 after 90 s (Fig. 4-c). The flex-DBD was operated at a voltage of 3 kV and discharge power of 0.5 W/cm$^2$, and the grounded surface temperature was T<50°C. We repeated the inactivation of *E. coli* using the Standard *E. coli* strain AMC 198 (ATCC 11229) (Fig. 4-d). Two experiments were conducted, one using a lower voltage, 2 kV peak voltage and the temperature of the grounded surface T<40°C, and 3 kV peak voltage and T<50°C. The higher applied voltage resulted in faster (p=0.003, ANOVA) inactivation of *E. coli*; $\log_{10}(N_o/N)$ = 5.8 after 180 s treatment (Fig. 4-d) as compared to $\log_{10}(N_o/N)$ = 2.6 after 180 s, demonstrating a dependence on the flex-DBD peak voltage. Temperature as a factor was eliminated in a separate experiment (below).

We calculated the $1\log_{10}$ reduction (D value) for *E. Coli* AMS 198 because the data is less variable than that of 10-beta, probably due to a greater control of the strain characteristics. At the start of the plasma treatment, the plasma affects the most susceptible bacteria that is located the closest to the plasma and hence is subjected to shorter-lived reactive plasma species. Hence the inactivation rate is the highest for the short treatment times. A linear fit to the treatment times of 10 s to 270 s give the times for $1\log_{10}$ reduction, D=74±2 s with the correlation coefficient, R=0.996. The high correlation coefficient also indicates a consistent operation of the flex-DBD device and its electrical characteristics and the temperature at the ground electrode were monitored throughout all experiments.

We tested whether it has been the device operating temperature (40°C and 50°C) that caused disinfection of *E. coli* 10 Beta and *E. coli* AMC 198 strains. Instead of treatment with flex-DBD we incubated contaminated coverslips at 50°C. We found no reduction in the bacterial load even at the longest exposure times of 180 sec and 270 sec. Therefore, the improvement in the disinfection at higher voltage may be attributed to plasma related effects such as the increase in the concentration of reactive species.

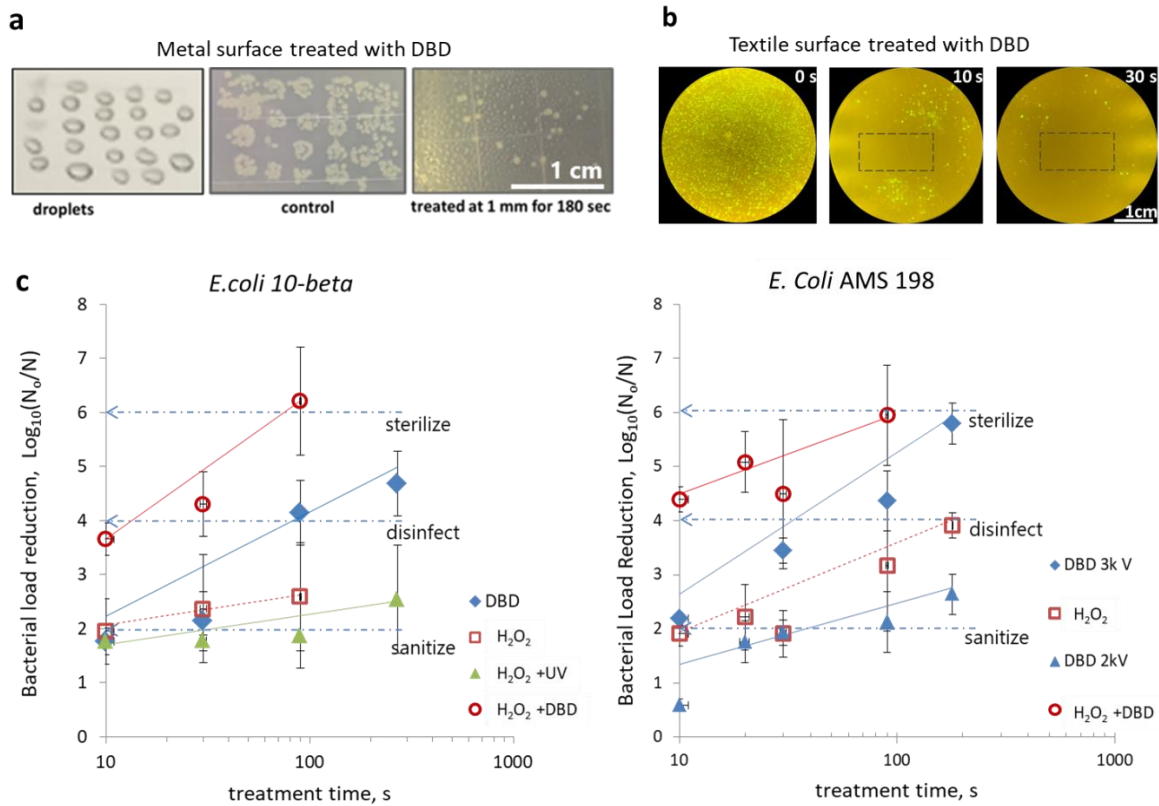

**Figure 4. Flex-DBD device effectively reduces bacterial load and its effect is synergistic with hydrogen peroxide.** (a) Aluminum surface inoculated with droplets of *E. coli* culture and stamped ("treated") photo). (b) Textile surface inoculated with *E. coli* (OP50-GFP) and stamped. (c) The reduction in the concentration of the *E. coli* colony forming units for two strains of *E. coli*. $\log_{10}(N_o/N)$, where $N_o$ is the number of CFU/ml surviving in the untreated samples, N is the number of surviving CFU/ml that remain after the treatment for 10 s, 30 s, 90 s, and *270* s. Common scale bars in a and b are 1 cm. The error bars are one standard deviation.

**Combined effect of plasma and hydrogen peroxide**

To augment the disinfection we applied the flex-DBD treatment in combination with 3% (by weight) $H_2O_2$ solution commonly available and, widely used for oral, skin, and wound disinfection. Applying the DBD together with $H_2O_2$ results in a 3.5 $\log_{10}$ reduction in the bacterial population in just 10 seconds and in >5$\log_{10}$ reduction in 90 seconds (Fig. 4-c and 4-d). The combined disinfection effect of hydrogen peroxide and plasma is faster than either DBD alone (p=0.04, paired ANOVA test) or $H_2O_2$ alone (p=0.03).

This synergy can be explained by the antibacterial mechanism of $H_2O_2$ solutions, which is based on the production of the highly reactive hydroxyl radicals. Hydroxyl radical is the highest oxidizer and it reacts with lipids in the cell membrane and oxidizes proteins and nucleic acids inside the cells. Because of its reactivity, it is very short-lived and needs to be produced at the site of action. Inside a cell, hydroxyl radical can be produced by the Fenton reaction:

$Fe^{2+} + H_2O_2 \rightarrow Fe^{3+} + OH^- + OH\bullet$ (2)

In addition, hydroxyl radical can be produced from the interaction of the superoxide ($O_2\bullet^-$) radical and $H_2O_2$:

$O_2\bullet^- + H_2O_2 \rightarrow O_2 + OH^- + OH\bullet$ (3)

Overall, $H_2O_2$ leads to the oxidation and eventually to cell death. The bacterial load reduction due to the $H_2O_2$ alone (Fig. 4-c and 4-d) slows down faster than the corresponding DBD treatment. Unlike the plasma generated reactive species, $H_2O_2$ is not replenished, and its concentration decreases with treatment time, which contributes to the decrease in the rate of bacterial inactivation. Linear fits the times for $1\log_{10}$ reduction, D=139±5 s, R=0.97 for *E. coli* 10-beta and D=135 ±5 s, R=0.947 *E. coli* AMC 198.

The rate of reduction increases to D=90 s, R=0.86 when the treatment is carried out by both the DBD and the $H_2O_2$ solution (Fig. 4-c). In humid environments, air plasma produces hydroxyl radicals generally through the interactions of electrons and excited nitrogen with water vapor. But the hydroxyl radical reacts, oxidizes or recombines to form hydrogen peroxide on a microsecond scale [38-40], which then diffuses into the solution. Hence plasma enhances the action of hydrogen peroxide as well as increases its concentration in the solution.

The results of the indicator experiments support the increase of $H_2O_2$ concentration in both the Luria Broth used for the *E. coli* suspensions and in the $H_2O_2$ solution used in the experiments with flex-DBD and $H_2O_2$. The application of flex-DBD to Luria Broth for 90 s increases the concentration of $H_2O_2$ (and other oxidizing agents) to at least 100 ppm. Applying the flex-DBD directly to the $H_2O_2$ solution for 90 s treatment time, the concentration of increases the concentration of $H_2O_2$ from 3% before treatment to 5-10% after treatment.

Plasma can generate $H_2O_2$ in the solutions but the UV radiation from the plasma can also decompose the existing $H_2O_2$. The experiments with the UV filter checked if the UV radiation from the flex-DBD was sufficient to explain the improved *E. coli* inactivation with $H_2O_2$. The results of blocking all plasma products except for the UV radiation improves the inactivation of bacteria in the first 30 s compared with $H_2O_2$ alone, but it does not achieve any additional reduction with increasing treatment time. After the first 30 s, the survival curve flattens (D>250 s). This effect of UV radiation is insufficient to explain the improvement in the inactivation with $H_2O_2$ achieved by the addition of the DBD plasma. Hence the plasma, not the UV alone, improves the action of $H_2O_2$.

Although many years of investigation continue to demonstrate the potential of plasma in disinfection of biological and non-biological surfaces, the devices that have been accepted alongside the steam sterilization in medical and pharmaceutical industries are low pressure plasma-$H_2O_2$ vapor chambers [28, 29]. This means that the most important role of plasma disinfection is not as a replacement for the standard bulk sterilization methods but in special niche applications and in personal consumer use [13].

Plasma devices such as the flexible-DBD device are portable and safe for consumer use that have the potential to provide disinfection without any chemical or biological residue.

**Conclusions**

Here we demonstrate the effectiveness of the first hand-held flex-DBD device that is suitable for personal use by untrained personnel. This flex-DBD device achieves fast disinfection, 4$\log_{10}$ reduction, in under 90 s and reaches >5$\log_{10}$ in 270 s. Augmented with $H_2O_2$ flex-DBD it achieves high bacterial load reduction two times faster than alone: >6$\log_{10}$ in 90 s with D=2.5 s for the first 10 s and D=30 s for 10 s – 180 s treatment times. These results are faster than UV and chemicals alone and faster than atmospheric pressure glow discharge, mesh DBD, and other plasma sources.

The flex-DBD device can make effective disinfection accessible to the untrained public. It is safe in operation since the user facing components are grounded. The electrical measurements described above show that the flex-DBD operates consistently over prolonged periods of time, at least 15 minutes at a time for a total of >100 hours of operation. The synergistic action with $H_2O_2$ reaches high levels of disinfection and opens exciting new possibilities for decontamination and treatment. This device can be used to disinfect personal protection equipment, hands, and frequently touched surfaces, as well as for wound treatment and other medical applications.

**Acknowledgements**

This material is based upon work supported by the U.S. Department of Energy, Office of Science, Office of Fusion Energy Sciences under contract number DE-AC02-09CH11466.
This research used resources of the Princeton Collaborative Low Temperature Plasma Research Facility (PCRF http://pcrf.pppl.gov ), which is a collaborative research facility supported by the U.S. Department of Energy, Office of Science, Office of Fusion Energy Sciences.

**Data availability statement:**

The datasets generated during and/or analyzed during the current study are available from the corresponding author on reasonable request.

**Author contributions statements**

SG collaborated on all experiments, data acquisition, analysis, writing the manuscript, and preparing the figures.

MBH collaborated on all biological experiments, conducted data acquisition for the biological experiments, and collaborated on writing the manuscript.

SY contributed the device design and main ideas for the experiments, manufactured the first prototypes of the device.

YR contributed the main scientific direction and guidance of this work, and collaborated on writing the manuscript.

GH contributed the biological expertise and guidance for the project, collaborated on writing the manuscript. GH is the PI under the PCRF collaboration with PPPL.

PE contributed scientific ideas for the experiments in the area of using the semipermeable material for the delivery of $H_2O_2$, collaborated on developing the scientific direction for this work.

**Competing interests**

The authors declare no competing interests.